# Crystallization Kinetics of Colloidal Spheres under Stationary Shear Flow

P. Holmqvist, M.P. Lettinga, J. Buitenhuis and Jan K.G. Dhont

Institut für Festkörperforschung, Forschungszentrum Jülich, D-52425 Jülich, Germany

**Abstract**
A systematic experimental study of dispersions of charged colloidal spheres is presented on the effect of steady shear flow on nucleation and crystal-growth rates. In addition, the non-equilibrium phase diagram as far as the melting line is concerned is measured. Shear flow is found to strongly affect induction times, crystal growth rates and the location of the melting line. The main findings are that (i) the crystal growth rate for a given concentration exhibits a maximum as a function of the shear rate, (ii) contrary to the monotonous increase of the growth rate with increasing concentration in the absence of flow, a maximum of the crystal growth rate as a function of concentration is observed for sheared systems, and (iii) the induction time for a given concentration exhibits a maximum as a function of the shear rate. These findings will be partly explained on a qualitative level.

## I. Introduction

Nucleation and crystal growth kinetics of suspensions containing spherical colloids has been studied the last decade by means of light scattering, confocal microscopy and by computer simulations. The induction time for nucleation, the number density of nuclei and the growth rate of crystals have been found to vary with concentration in a way that depends on whether hard spheres or charged spheres are used. A number of the observed phenomena have been explained on the basis of an extension of classical nucleation theory to colloids.[1] So far, no experiments have been reported where the effect of flow on the kinetics of nucleation and crystal growth is considered and where the shear-rate dependence of phase transition lines has been measured. On applying flow, crystal growth rates are not just determined by diffusion of spheres in the liquid phase towards crystal interfaces, but also by convective mass transport, both from particles in the liquid to the crystal interface as well as particles that are sheared off the interface into the liquid. The latter is usually referred to as erosion. In addition, the probability for density fluctuations giving rise to stable nuclei will be affected by flow. *The present paper reports on nucleation and crystal growth rates in shear flow as well as the shear-rate* *dependent location of the melting line for a dispersion of charged colloidal spheres.* Besides the concentration, the shear rate is now an additional variable. As far as we know the only experimental paper on crystal growth under shear flow is by Tsuchida [2], who found a small decrease in the growth rate with increasing shear rate. Shear flow is found in the present paper to strongly affect induction times, crystal growth rates and the location of the melting line.

The theory of crystal growth kinetics in the absence of flow has been developed on the basis of semi-phenomenological equations of motion, where driving forces are often formulated in terms of thermodynamic quantities.[3-5] There is no theoretical approach on this level formulated yet that describes crystallization kinetics under flow conditions. A first simulation study on this subject has been published recently, the predictions of which, however, can not be directly related to the quantities of experimental importance in the present paper.[6]

Experiments on the response of single crystals to flow have been performed for the first time by Hoffman[7] and later by Ackerson et al.[8,9] The aim of that work was to study the microstructural response of crystals under (oscillatory) flow. These experiments reveal flow alignment of single crystals in flow, but do not consider the kinetics of nucleation and crystal growth under flow conditions. In accordance with these earlier experiments, in the present paper we also find flow alignment of crystals once they are large enough.

The study of nucleation and crystal growth of colloids probably contributes to our understanding of simple molecular systems as well.[10,11] The advantage of colloidal systems as compared to molecular liquids is that they are experimentally more easily accessible because of the much larger time- and length scales involved. Moreover, the inter-colloid particle potential can be varied from steeply attractive to long ranged repulsive, leading to very different types of phase transitions and non-equilibrium states like gels and aggregates. When the interaction potential is engineered to be hard sphere like, heterogeneous and homogeneous nucleation is observed.[12] At high concentrations, hard-sphere systems get trapped in a glass state and are thus not able to reach the equilibrium crystalline state. The glass can be made crystalline by imposing an (oscillatory) shear flow[13] When the interaction potential is engineered to be long-ranged repulsive, making the spheres highly charged, crystallization is observed at much lower concentrations. Due to the low particle concentration these systems exhibit a very low yield stress as compared to hard spheres. The charged system studied in the present paper, like other charged systems,[14] exhibit a re-entrant crystallization behaviour, where the crystals melt at sufficiently high concentrations so that the fluid state is stable instead of the glass state as formed by hard spheres. At even higher concentrations, where hard-core interactions become dominant, crystallization should occur again, and a glass is probably formed at even higher concentrations like for hard spheres. At lower concentrations there are thus two melting points in the absence of shear flow. In the shear-rate versus concentration diagram the melting line is a closed curve, the location of which is determined in the present paper by means of light scattering and microscopy.
Nucleation and crystal growth kinetics are characterized by the induction time, the number density of nuclei and the growth rate of crystals. Despite the large interest in this field during the last decade, there are still open questions concerning crystallization kinetics even in the absence of flow. It has been observed, for example, that the growth rate as a function of concentration has a distinct maximum in some cases[15-17] and shows a constant increase in other cases.[18, 19] The maximum is predicted from simulation,[3] while a constant increase is predicted from the classical theory by Wilson and Frenkel.[4, 5] A population balance model of the growth kinetics, which takes into account the constant decrease of the number of particles in the liquid phase, shows good agreement with the observation of a maximum as a function of concentration.[20] For colloids, the occurrence of a maximum in the crystal growth rate might be explained by means of a reduction of the diffusion coefficient with increasing concentration due to hydrodynamic interactions between the colloidal spheres.[1] No theoretical prediction exists yet for the concentration dependence of the induction time, but there seems to be a consensus in how to determine it. Two different concentration dependences of the induction time have been observed: one where the induction time monotonically decreases with increasing concentration[16-19] and another which shows a minimum at a specific concentration.[15, 17, 20]

Very little is reported on the actual kinetics of the crystallization under shear flow.[21] Most information on crystallization kinetics has been obtained by simulations.[6, 22, 23] According to Butler and Harrowell[6], there are three main effects of steady shear on crystallization kinetics : (i) Crystallites will orient in shear flow, otherwise they will be destroyed, (ii) the crystallite will be destroyed due to convection if crystal layers move past one another faster than they grow, which is referred to as erosion, (iii) the nucleation and growth rate depends on the chemical potential difference between liquid and crystal state, which difference is affected by flow. Which of these processes dominate depends on the shear rate and concentration.

The experimental system used in the present study is TPM-coated silica in a mixture of toluene and ethanol.[24] The colloidal spheres are charged and interact through their double layer repulsion. This system has been used before to study nucleation and crystal growth kinetics in the absence of flow.[16] The interaction potential is less repulsive as compared to aqueous systems, where the ionic strength is controlled using ion-exchangers [25]. On the other hand, the potential is significantly affected by the charges on the particles, resulting in a very different behaviour as compared to hard spheres. In our experiments we use Small-Angle Light Scattering (SALS) under shear to follow the kinetics of nucleation and crystal growth. In addition, heterodyne dynamic light scattering is employed to measure velocity profiles in the polycrystalline samples. Data are partly interpreted in view of the work by Butler and Harrowell[6] and the findings of Dixit et al..[6, 20]

## II. Experimental

### II.1. Small Angle Light Scattering (SALS)
We used a home-built optical couette shear cell combined with a SALS set-up. The shear-cell consisted of a rotating inner cylinder with a diameter of 43 mm and a static outer cylinder with a diameter of 47 mm, resulting in a gap width of 2 mm. The inner and outer cylinders are both made of optical grade glass. A 10 mW diode pumped crystallaser (Laser 2000) with a wavelength of 440 nm was used as a light source. In order to have the laser beam going through the gap just once it is directed through the centre of the rotational axis of the inner cylinder. In the rotating cylinder the beam is directed along the gradient direction with a prism so that the flow–vorticity plane is probed (see Figure 1). Scattered intensities are projected on a white semi-transparent screen and images were taken in transmission mode with a Peltier cooled 12-bit CCD camera, with 582 × 782 pixels (Princeton Instruments, microMAX). The scattering angle of the first order Bragg peak was 30º, which corresponds to a wave vector q of 0.0108 nm$^{-1}$. The shear rate was varied between zero and $\dot{\gamma}=1$ s$^{-1}$.

### II.2. Velocity profiles

In principle, velocity profiles might be non-linear during crystal growth, in which case the shear rate is not a constant throughout the gap of the shear cell. In order to verify whether the flow profile remains linear during crystal growth, spatially resolved heterodyne dynamic light scattering experiments have been performed. The setup is based on a differential Laser Doppler velocimeter[26] using a 1/1 beam splitter (Spectra Physics, Krypton laser λ=637 nm). The two beams are then focused on the same spot in the gap of the optical couette cell. The intensity auto correlation function of the scattered light from the overlapping region in the direction parallel to the incident beams exhibits an oscillatory component, the period of which is inversely proportional to the local flow velocity at the point where the two laser beams intersect.[27] The characteristic frequency f of the oscillation is equal to,

$$f = \frac{2\sin(\theta/2)n}{\lambda}v$$

where θ is the angle between crossing beams (21º in our set up) and n=1.465 is the refractive index of the dispersion and s is the local suspension velocity.

### II.3. Synthesis of the TPM-silica particles

The colloidal particles used in this study are 3-(Trimethoxysilyl)propyl methacrylate (TPM) coated silica particles. These particles are charged and can be dispersed in an ethanol/toluene mixture with up to 80% toluene.[24] Silica core particles with about half the final size were synthesized according to Stöber.[28] These particles were grown to there final size by continuous addition of a TES/ethanol mixture as described in[29] and then coated with TPM according to Philipse et al..[24]. The particles were purified from unreacted reagents by repeated cycles in which the particles are sedimented by centrifugation and redispersed in ethanol. Dynamic light scattering analysis of a dilute dispersion yields a hydrodynamic radius of the particles of 202 nm. Image analysis of transmission electron microscopy photographs gives an average radius of 206 nm with a relative standard deviation of the size distribution of 6%.

### II.4. Experimental Results

The concentration range where homogeneous nucleation occurs was determined by visual inspection. The concentration range where homogeneous nucleation is observed within 24 hours after homogenization is 25.0±0.3 and 29.0±0.3 wt % of silica. It is in this region that the crystallization kinetics was investigated both as a function of shear rate and concentration. The concentration region where crystallization is observed is substantially lower than what is expected for hard spheres, but not as low as found for de-ionized aqueous systems.[25] The charge of the colloids estimated from the location of the crystallization region is about 300 elementary charges per colloidal sphere, which complies with the charge density as reported for this system in a toluene/ethanol mixture by Philipse et al..[24] Due to the low degree of dissociation of the counter ions in the organic medium, the ionic strength in the solution is very low (less than 10 μM), resulting in a large Debye length (about 100 nm, which is half the radius of the silica spheres). This leads to soft interactions and a reasonable large concentration window to investigate crystallization kinetics. The phases surrounding the crystal region are found to be fluid-like and do not display any Bragg reflections. On increasing the concentration of silica particles the crystals melt, probably as a result of either the very long-ranged character of the interaction potential (which is the reason that star-like polymers exhibit re-entrant crystallization behaviour)[30-32] or of the change of the interaction potential between the particles due to change of their charge and/or of the Debye length as the concentration is changed. This has been observed in other types of suspensions of charged particles as well.[14]

The region where crystallization is investigation is bounded by the two above mentioned melting transition concentrations. Within this concentration range, the equilibrium state is a state where all colloidal particles are crystallized. The concentration range where crystals are in equilibrium with a colloidal fluid is very small in comparison with the region spanned by the two melting concentrations.

As can be seen from Figure 2, the intensity auto-correlation function at low concentrations, below the concentration where crystallization is observed, exhibits essentially a single-exponential decay, as usual for a concentrated fluid. For high concentrations, above the concentration where crystallization ceases to occur, the (ensemble averaged) auto-correlation function still decays to zero for long times, which shows the ergodic nature of the system, but there are now two distinct decay mechanisms. The system is ergodic but exhibits a two-step decay similar as for a glass. The interesting dynamical behaviour of this type of fluid is beyond the scope of the present paper and will not be discussed here further. The high density stable liquid state will be referred to hereafter as the liquid-G phase (where the G refers to the glass-like two-step decay of the density auto-correlation function typical for a glass). It is expected that at higher concentrations, where hard-core interactions become dominant, crystallization occurs again, followed by a glass transition at even higher concentrations.

When studying crystallization kinetics, one needs to start from a well-defined initial state. Therefore, we first pre-shear the samples at a shear rate of $\dot{\gamma}$=10 s$^{-1}$, which, as will be seen later, is well above the melting line of the investigated homogenous crystallization region for all concentrations. This high shear rate also destroys any previous history of the sample. Subsequently the shear rate is quenched down to zero or a finite shear rate where the crystallization behavior was then observed with SALS. The nuclei/crystals that are formed will show Bragg reflections when they are sufficiently large. The scattered intensity from these nuclei/crystals increases as they growth in size. After the shear rate quench the time evolution of the first Debye-Scherrer ring was monitored, taking three images per second. In Figure 3 the resulting Debye-Scherrer ring is presented for two different shear rates, $\dot{\gamma}=0$ s$^{-1}$ (a,b) and $\dot{\gamma}=0.05$ s$^{-1}$(c,d), and at two times, 0 s (a,c) and 1500 s (b,d).

Visual inspection of these scattering patterns reveals some important features of the crystallization process. At early times (a,c) no Bragg scattering can be observed and only the typical structure factor maximum of the meta-stable fluid state can be seen as an intensity ring This behavior is independent of the concentration and shear rate. The structure factor maximum is located at the same q-vector as the Bragg peaks for low concentrations, while for the two highest investigated concentrations the Bragg peaks shifts to higher q-values during crystallization. The change of the peak position indicates that the system at high concentration has a significantly smaller lattice constant as compared to crystals formed at lower overall concentration. The rate of this decrease depends on the concentration and shear rate. After the quench, but before the fluid scattering ring significantly decreased, Bragg reflections appear and disappear at random with time. After some time, most Bragg reflections do not disappear anymore and the number and lifetime of these reflections increases with time. For zero shear rate the Bragg reflections appear randomly over the Debye-Scherrer ring, as can be seen in Figure 3b. These Bragg reflections increase in intensity over the whole Debye-Scherrer ring while new reflections continue to form until the sample is completely filled with crystals. In the case of a low concentration and low shear rate this random Bragg peak distribution reflects the initial random orientation of the crystals. For all other shear rates and concentrations the Bragg peaks eventually form at six, well defined radial positions, as can be seen in Figure 3d. Apparently, shear flow orients all crystal structures in the same direction. This six-spot pattern, which is independent of shear rate and concentration, indicates a hexagonal packing along the flow direction. This hexagonal orientation of the crystals is similar to earlier reported studies on sheared single crystals.[9] Where the six-spot pattern is observed, the system is still partly fluid like. Hence, not only single crystals, but also crystals floating in a fluid orient in a preferred direction under flow, despite the fact that visual observation reveals that crystal geometries are not very anisotropic.

In order to quantify our data we plot the intensity of the first order Bragg reflections as a function of the rotational angle θ, shown in Figure 4. The integrated area of the Bragg peaks is considered, which is obtained after subtraction of the background and liquid structure factor peak intensity from the total measured intensity. θ=0º is defined as the flow direction and consequently θ=90º and θ=270º are along the vorticity direction. Examples for two different concentrations, 26.5 wt% (a,c) and 28.0 wt% (b,d), at two different shear rates, $\dot{\gamma}=0$ s$^{-1}$ (a,b) and $\dot{\gamma}=0.05$ s$^{-1}$(c,d), are plotted in Figure 4. From this plot we can infer the time evolution of the Bragg reflections over the entire Debye-Scherrer ring. During the very initial stages of phase separation, typically no Bragg peaks are detected. In some systems, however (e.g. a,c), temporary Bragg reflections can be found before a systematic growth occurs. As expected, there is no orientational order of the Bragg reflections in the absence of flow, i.e. the crystal regions are randomly orientated throughout the scattering volume at all times. At lower concentrations, relatively large crystals are observed throughout the sample. This is reflected in the scattering spectra as fewer first order Bragg reflections are seen due to the smaller probability of fulfilling the Bragg condition (see Figure 4a). For higher concentrations, the observed domains are smaller and their number density is larger as compared to lower concentrations. As a result, the probability to fulfil the Bragg condition for scattering into the observed scattering plane is larger, as is evident from Figure 4b. The effect of shear is also clear in these plots, as can be seen from Figure 4c,d. For the low concentrations (Figure 4c), the Bragg peaks appear randomly at first but eventually a hexagonal scattering pattern appears (Figure 4c). For the high shear rate (see Figure 4d), the hexagonal pattern is formed essentially immediately and does not change during the crystallization process.

From intensity profiles as shown in Figure 4, crystal growth rates and the induction times can be extracted as functions of both concentration and shear rate. In Figure 5 the time dependence of the total scattering intensity of the first order Bragg peaks is presented for a quench to $\dot{\gamma}=0$ s$^{-1}$ (a) and $\dot{\gamma}=0.10$ s$^{-1}$ (b) for various concentrations. Here, the total scattered intensity is the integral of the scattered intensity over all scattering angles of the profile plotted in Figure 4. The non-monotonic character of the curves is due to the dynamic process of disappearing and re-appearing of crystalline domains in the scattering volume. The growth rate is defined as the slope of the total intensity as a function of time (disregarding the fluctuations due to accidental Bragg reflections) and the induction time is defined as the intercept of the straight line with the time axis. The experimental determination of crystal growth rates and induction times is illustrated in Figure 5b. When the system is quenched to zero shear rate it is apparent that the induction time decrease and growth rate increase with concentration, which is in accordance with previous results on non-sheared systems.[15-19]

Growth rates and induction times, as obtained from the total intensity vs. time curves as described above, are given as a function of shear rate for different concentrations in Figures 6 and 7, respectively. The growth rate in Figure 6 exhibits a maximum as a function of shear rate. The location of the maximum shifts to lower shear rates with increasing concentrations (at the highest concentration the maximum is located at a shear rate that is lower than the minimum applied shear rate). Figure 7 shows that the induction time also displays a maximum as a function of shear rate, which decreases in height with increasing concentration. The concentration dependence of the growth rate (inset of Figure 6) for zero shear rate does not show a maximum, in contrast to earlier findings for a similar system.[16] Such a maximum is only found for somewhat higher shear rates. The induction time decreases continuously with increasing concentration for zero shear rate (see inset Figure 7), similar to what is found in ref.[16] At finite shear rates, however, a maximum in the induction time as a function of concentration is found.

On the basis of visual observation of the first Debye-Scherrer ring and the analysis of the growth rate and induction time as discussed above, we are able to construct a non-equilibrium phase diagram, that is, we can determine the location of the melting line in the shear rate versus concentration plane. The phase boundaries were defined as the concentration and shear rate where no Bragg reflections are observed within 60 minutes after the shear quench. We determine the phase boundaries by this visual observation as well as by using the plots in Figures 6,7 of the growth rate and induction time. From these plots the phase boundary is determined from the shear rate and concentration where the growth rate and induction time are extrapolated to zero. The resulting non-equilibrium-phase diagram of the homogeneous crystallization region is shown in Figure 8. The error bars indicate the spread in the location of melting points as obtained from the three methods mentioned above.

In order to interpret the data discussed above, it is important to know whether the shear rate is a constant throughout the gap during crystallization. Flow profiles during the crystallization were measured for the 25.2 wt % sample by means of heterodyne light scattering, for a gap width of 2 mm. Flow profiles were measured during one hour after a shear quench, for three different shear rates. The measurement of a flow profile takes about 5 minutes. Results are collected in Figure 9. Flow profiles collected for a particular shear rate do not change during the nucleation and crystallization process. All shear rates show a linear decay of the velocity starting 300 μm from the inner rotating wall (located at 2 mm in Figure 9). In the case of the highest shear rate, $\dot{\gamma}=0.15$ s$^{-1}$, the extrapolation of the linear profile to the inner wall intersects at 0±0.05 mm. For the two lowest shear rates, $\dot{\gamma}=0.05$ s$^{-1}$ and $\dot{\gamma}=0.10$ s$^{-1}$, this extrapolation intersects at 0.16±0.05 and 0.14±0.05 mm, respectively. These findings comply with stick boundary conditions, and are in accordance with observations done with a microscope at the outer wall, where stationary crystal domains where seen to form. These crystal domains grow until they get large enough to be caught up in the shear flow further out into the gap. The flow profiles are thus linear throughout the gap, except for a small region close to the walls. This renders the shear rate in the bulk essentially equal to the applied shear rate. No shear induced phenomenon like shear banding influences the nucleation and crystal growth rates as discussed above.

### III. Summary and Discussion
The observations described in the present paper are essentially concerned with the concentration and shear rate dependence of the crystal growth rates (Figure 6), the induction time (Figure 7) and the shear-rate dependent location of the melting line (Figure 8). In the following we shall speculate on possible mechanisms that could explain our experimental findings. So far there are limited analytic theories and simulation work done on crystallization kinetics under shear. Note that the melting line plotted in Figure 8 for low concentrations is almost vertical. The effect of flow on the location of the transition is therefore small for these small concentrations, implying that the chemical potential difference Δμ between the liquid and crystal is almost independent of the shear rate. The effects of flow at low concentrations are therefore predominantly due to its effect on mass transport rather than the driving force Δμ. At intermediate and high concentrations, however, flow may also affect the driving force Δμ. At high concentrations, where Δμ is large, the effect of flow is relatively small as compared to low concentrations.

The effect of convection is that it enhances crystallization rates at low shear rates, since spheres from the liquid are convected to the crystal. Slowing down of growth rates due to depletion of spheres as a result of diffusion as predicted by Dixit[20] no longer applies to sheared systems. At high shear rates convection is destructive since particles are now sheared off the crystal surface into the liquid (erosion), while particles in the liquid phase have no time to be incorporated into the crystal structure. This explains why *the crystal growth rate for a given concentration exhibits a maximum as a function of the shear rate*. For high concentrations, where the chemical potential difference $\Delta\mu$ between the liquid and crystal is large, the growth rate is large, which diminishes the relative effect of convection. This is why the maximum growth rate occurs at lower shear rates for higher concentrations.

The crystal growth rate (see Figure 6) shows a marked, monotonous increase as a function of concentration. *Contrary to this monotonous increase of the growth rate with increasing concentration in the absence of flow, a maximum of the crystal growth rate as a function of concentration is observed for sheared systems.* This maximum is in accordance with the population balance model[20].

The orientation of crystals can have an additional effect on the growth rate. The sheared liquid probably forms strings of colloidal spheres similar to the sheared crystal structure.[9] This might effectively reduce the surface tension and enhance the growth rate. In addition, when applying shear flow, only those nuclei will survive and grow, which will orient along the flow direction sufficiently rapidly. If they do not orient to the flow sufficiently rapidly, they will be destroyed due to strain forces.

The effect of flow on nucleation rates and induction times is much more difficult to understand as compared to crystal growth kinetics. Here, the effect of flow on the probability of fluctuations that lead to the formation of stable nuclei must be explained. Microstructural order of the meta-stable liquid will be affected by flow, which is one reason for the shear-rate dependence of the probability for formation of stable nuclei. We find that *the induction time exhibits a maximum as a function of the shear rate* (see Figure 7). The increase of the induction time with increasing shear rate for the lower rates is probably connected to the suppression of nuclei formation as found in simulations[22]. The decrease, at higher shear rates, probably has the same origin as for the crystal growth rate related to orientation and surface tension.

At high concentrations, the large number of crystallites causes an almost instantaneous collective reorientation, even at small shear rates, and no effect of flow on the induction time is expected, as far as its disruption of nuclei due to non-oriented nuclei is concerned. Note that the magnitude of the induction time is very small for high concentrations. Like for the growth rate, the relative effect of flow on nucleation times at high concentrations is relatively unimportant due to the large chemical potential difference $\Delta\mu$ that drives the formation of nuclei.

Without flow, the driving force for crystallization is the chemical potential difference between the super cooled liquid and the crystalline phase, while the rate limiting process is diffusion. In the above qualitative discussion we have used the notion of a chemical potential in flow to interpret our findings. However, in the presence of flow, it is in principle not possible to define a chemical potential, since shear flow is a non-conservative external field. It might nevertheless be that formal definitions of a chemical potential in sheared systems may describe the essential features of the driving force for nucleation and growth also for systems under shear flow. Here, the shear-induced structural deformation of the liquid state and orientation of nuclei and crystals may be of importance. Apart from diffusion, in a sheared system also convection plays an important role for the kinetics of crystal growth. This problem has been addressed in simulations,[6] but not to an extent that it allows an unambiguous comparison with the present experiments.

**Acknowledgement**

We thank Hartmut Kriegs for his assistance on heterodyne dynamic light scattering. P. H. was supported by the EU project "Hard to ultra soft colloids". This work has been supported by the Transregio SFB TR6, "Physics of colloidal dispersions in external fields".

**Figure 1.** Schematic representation of the SALS-Rheology setup (upper figure) and the top view of the shear cell (lower figure).

**Figure 2.** Correlation functions, $C(q,t)$, at $q=0.0075$ nm$^{-1}$ for two concentration 20 wt% (below crystallization region) and 35 wt% (above the crystallization region)

**Figure 3.** The first Debye-Scherrer ring for two different shear rates, $\dot{\gamma}=0$ s$^{-1}$ (a,b) and $\dot{\gamma}=0.05$ s$^{-1}$(c,d), at two times, 0 s (a,c) and 1500 s (b,d).

**Figure 4.** The radial distribution of the Bragg reflection intensity of the first Debye-Scherrer ring for two different shear rates, $\dot{\gamma}=0$ s$^{-1}$ (a,b) and $\dot{\gamma}=0.05$ s$^{-1}$(c,d) of two different concentrations, 26.5 wt% (a,c) and 28.0 wt% (b,d).

**Figure 5.** The time dependence of the total Bragg peak intensity of the first Debye-Scherrer ring for $\dot{\gamma}=0$ s$^{-1}$ (a) and $\dot{\gamma}=0.05$ s$^{-1}$ (b) at four different concentrations. The inset in (b) illustrates the determination of the growth rate and induction time.

**Figure 6.** Growth rate of colloidal crystallization under shear as a function of shear rate for four different concentrations. Inset: the concentration dependence of the growth rate for four different shear rates. All points are averages of two measurements and have an error of less than 20%

Figure 7. Induction time of colloidal crystallization under shear as a function of shear rate for four different concentrations. Inset: the concentration dependence of the Induction time for four different shear rates. All points are averages of two measurements and have an error of less than 20%

**Figure 8.** The non-equilibrium phase diagram of the homogenous crystallization region as a function of shear rate and concentration. The surrounding liquid phase are described in the text

**Figure 9.** Flow profiles for the 25.2 wt% samples at three different shear rates (□) 0.05 s$^{-1}$, (○) 0.10 s$^{-1}$ and (△) 0.15 s$^{-1}$.

Figure 1

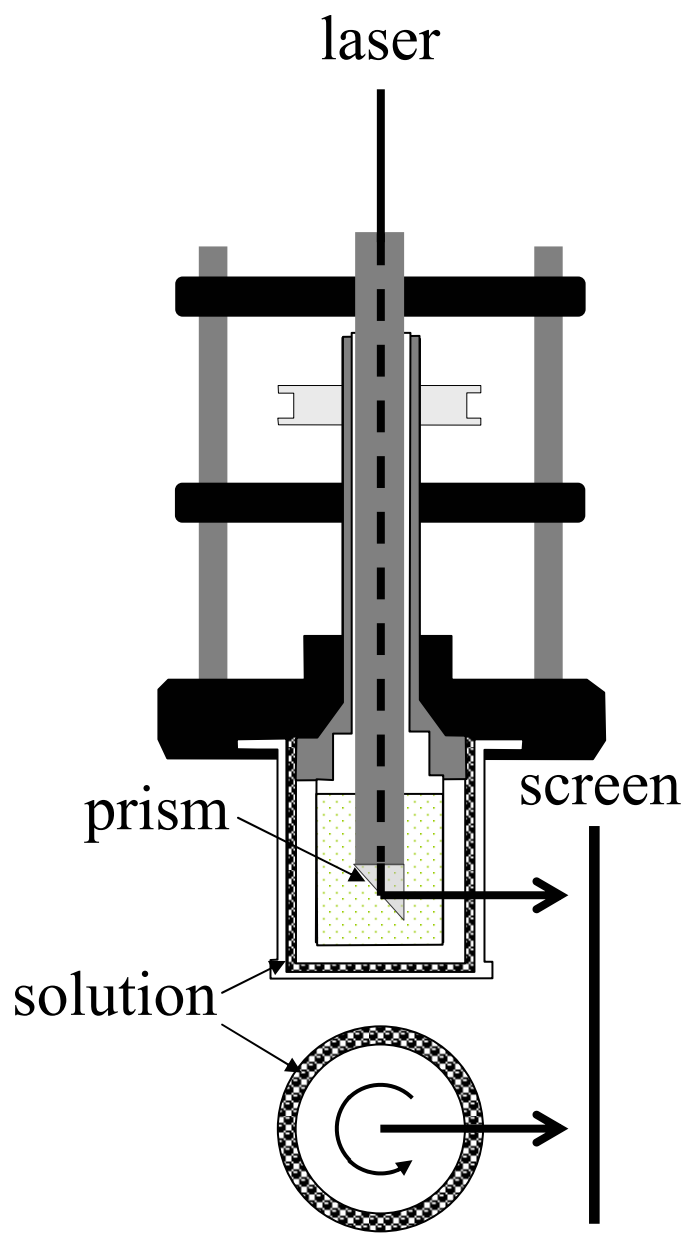

Figure 2

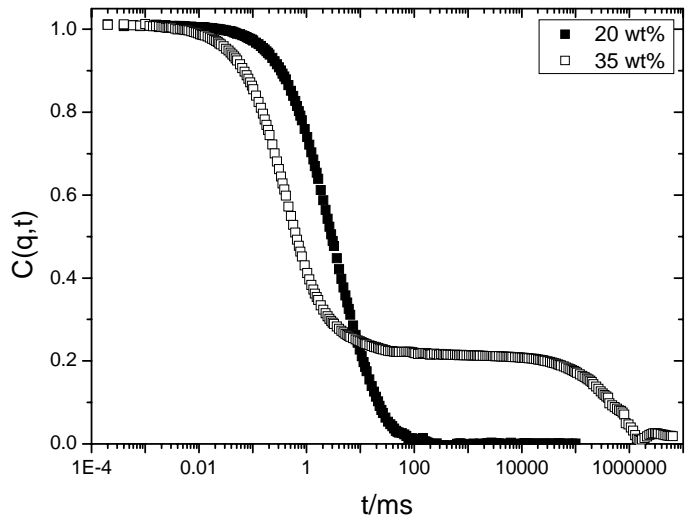

Figure 3

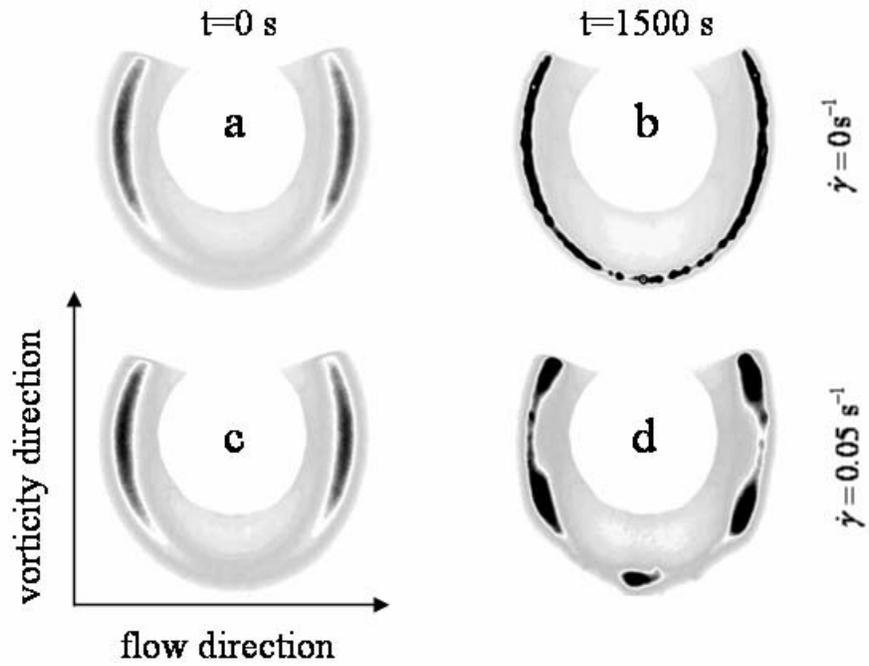

Figure 4

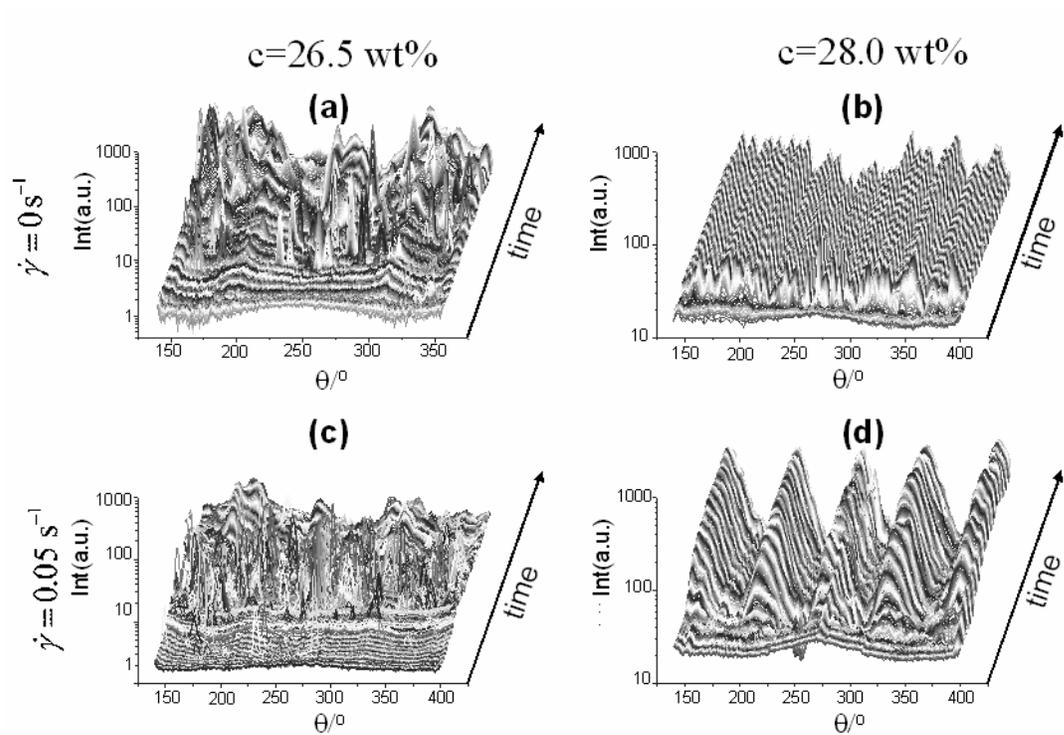

Figure 5a

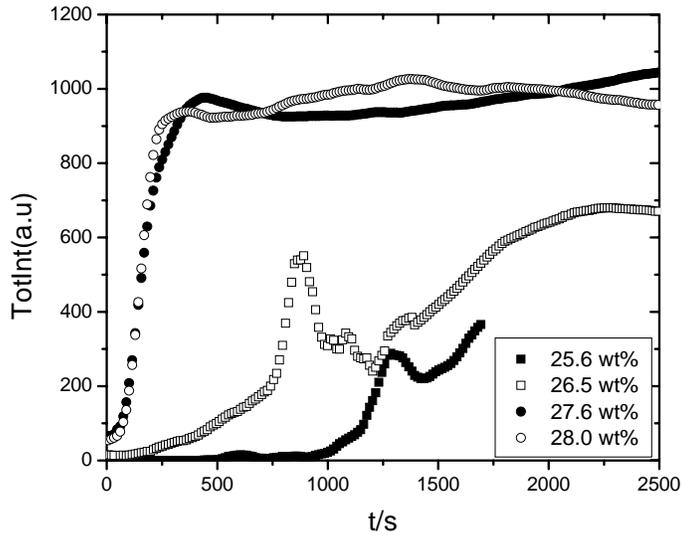

Figure 5b

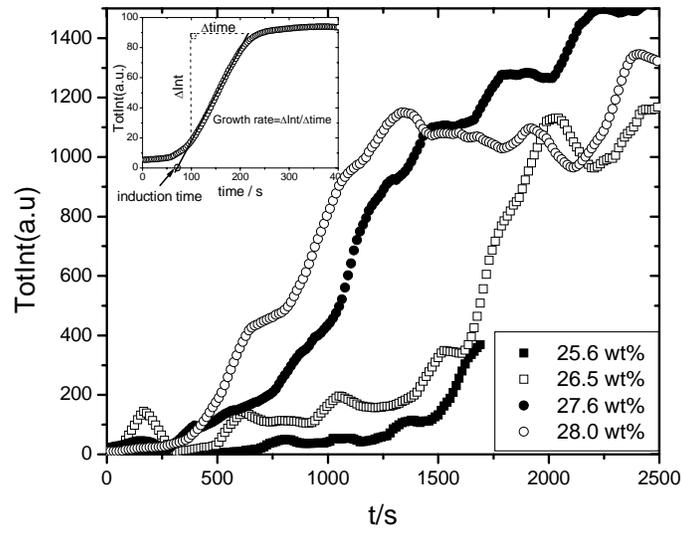

Figure 6

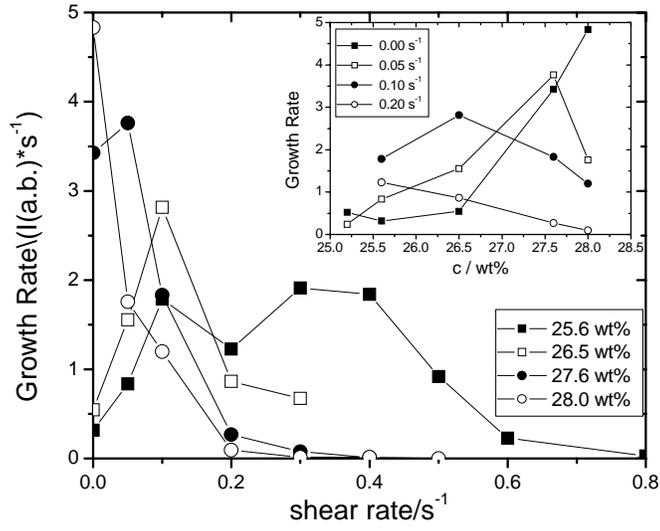

Figure 7

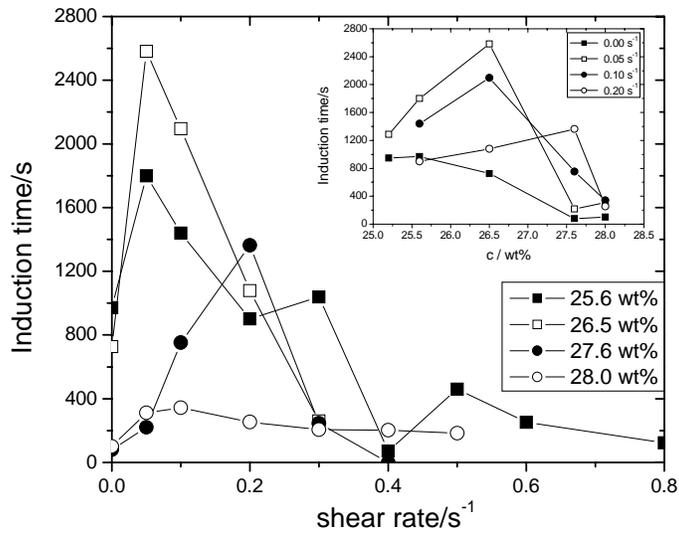

Figure 8

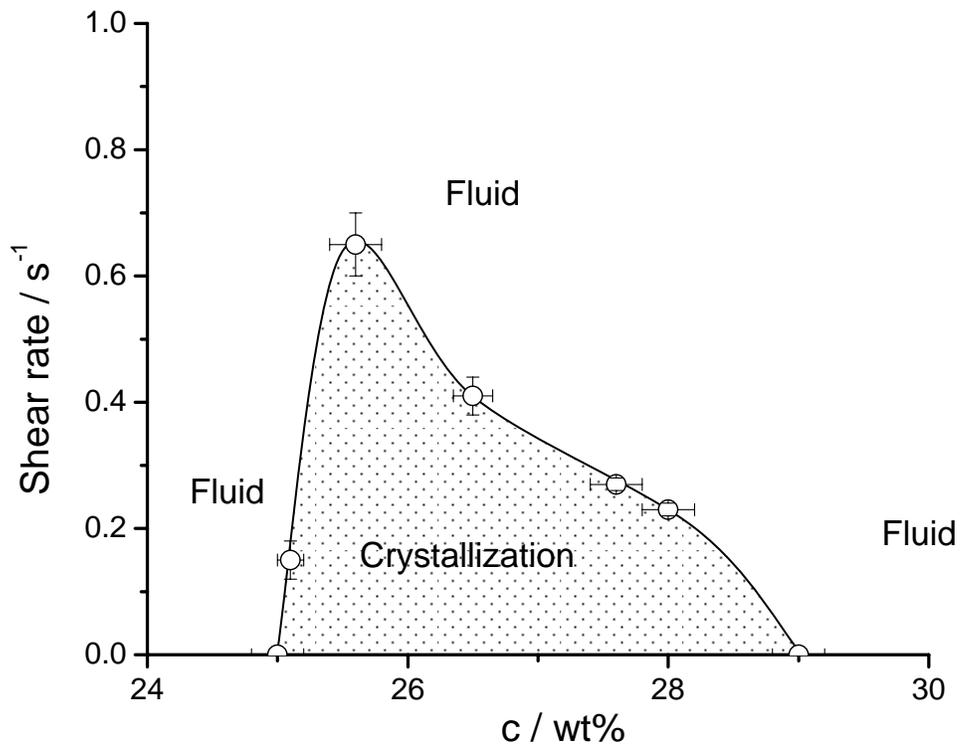

Figure 9

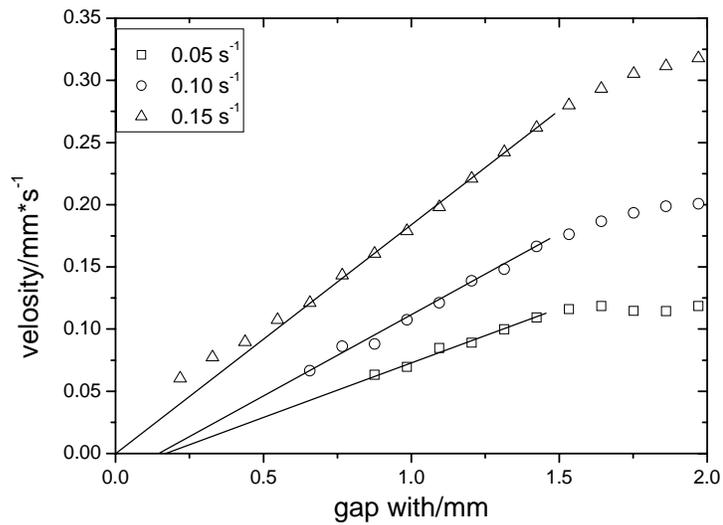